# How Many Long-Range Orders Are in the Abrikosov State

A.V. Nikulov

*Institute of Microelectronics Technology, RAS, 142432 Chernogolovka, Moscow Region, Russia.*

**Abstract.** It is argued that only a single spontaneous long-range order, namely the phase coherence exists in the Abrikosov state, and the prediction of the crystalline long-range order of vortex lattice does not correspond to the facts.

**Keywords:** Abrikosov vortex state, thermal fluctuations, vortex lattice melting.
**PACS:** 74.25.Qt, 74.25.Op

## INTRODUCTION

Superconductivity is one of the best known macroscopic quantum phenomena. The first experimental evidence of this wonderful fact is the Meissner effect observed in 1933, and the most marvelous one is the Abrikosov vortex state [1]. A superconductor expels the magnetic flux $\oint_l dl A = \Phi$ when the macroscopic wave function $\Psi = |\Psi|exp(i\varphi)$ does not have singularities, and the integral $\oint_l dl \nabla \varphi$ of the gradient $\nabla\varphi = (mv + 2eA)2\pi/h$ of its phase $\varphi$ along any closed path l within superconductor should be equal zero: $\oint_l dl \nabla \varphi = 2e\Phi 2\pi / h = 0$. When, however, the phase $\varphi$ of the wave function $\Psi$ has singularities inside *l,* its circulation equals

$$\oint_l dl \nabla \varphi = 2\pi n \qquad (1)$$

where *n* is the number of singularities, and consequently $\Phi = hn/2e = n\Phi_0$ Therefore, numerous direct observations of the proportionality $\Phi = n\Phi_0$ between the magnetic flux $\Phi$ and the number *n* of the Abrikosov vortices prove that the Abrikosov state is a mixed state possessing a long-range phase coherence.

## PREDICTION OF THE VORTEX LATTICE AND EXPERIMENTAL DATA

The attention of most people was attracted to the additional spontaneous crystalline long-range order predicted by the famous work of Abrikosov [1] and others [2]. Observations of the ordered vortex lattices created an illusion that theoretical prediction of a vortex lattice with a long-range crystalline order received an experimental corroboration.

## Spontaneous or Constrained Order

One should, however, realize the fundamental difference between the spontaneous long-range order predicted by [1,2] in homogeneous infinite symmetric space and an order observed in non-homogeneous asymmetric space of a real superconductor sample of finite sizes.

No observation can corroborate the prediction [1,2] of the spontaneous crystalline long-range order of vortex lattice, since it cannot be realized in any real sample with pinning disorders [3]. Moreover, many experimental results are an evidence of non-spontaneous nature of this order constrained by the asymmetry of the underlying atomic lattice. The question about the possibility of the Abrikosov state in the ideal case of the homogeneous infinite symmetric space is important first of all for numerous theories of the vortex lattice melting, since this phase transition should be connected with spontaneous, unconstrained long-range order.

## Two Long-Range Order and a Single Phase Transition

There is a discrepancy between the prediction of two long-range order predicted in [1,2] and only a single second order phase transition assumed to occur at the second critical field $H_{c2}$. The situation became more dramatic in the seventies, when the consideration of the thermal fluctuations showed that this transition, assumed during a long time, can not exist because of the reduction of the effective dimensionality of thermal fluctuations on two near $H_{c2}$ [4]: two long-range orders but no transition. The lost transition into the Abrikosov state was found first in bulk superconductors at $H_{c4} < H_{c2}$ in the early eighties [5]. This result was repeated in ten years on high-Tc superconductors (HTSC) [6] and it was shown that this phase transition is first order [7].

## VORTEX LATTICE MELTING OR DISAPPEARANCE OF THE PHASE COHERENCE

Thus, according to all experimental results [5,6] only one phase transition is observed at $H_{c4} < H_{c2}$. The interpretation of this transition as vortex lattice melting became popular, since the Abrikosov state is first of all the vortex lattice for most people. Most naive theorists used the Lindemann criterion for description of this transition [8,9]. But some experts realize that the Abrikosov state is first of all the mixed state with long-range phase coherence [10]. Nevertheless, they do not reject the concept of vortex lattice melting. The reason of such an attitude is the definition of phase coherence used by all theorists.

### Definitions of Phase Coherence in the Abrikosov State

According to this definition using a correlation function, i.e. when coherence between two points is considered, the long-range phase coherence can exist in the Abrikosov state described by multi-connected wave function only if this function has a periodical long-range order. In this case the vortex lattice melting is simultaneously the disappearance of long-range phase coherence and the concept of vortex lattice melting should not be rejected. But the definition by the use of a correlation function comes into conflict with the experimental evidence $\Phi = n\Phi_0$ of the phase coherence (1) according to which the existence of the long-range phase coherence can not depend on an arrangement of vortices inside a closed path $l$. The observation of the vortices at any arrangement is evidence of phase coherence since singularities in the mixed state with phase coherence can not exist without phase coherence.

### The Abrikosov State Is Not Vortex Lattice with Spontaneous Long-Range Order

The comparison [11] of the position $H_{c4}$ of the transition into the Abrikosov state experimentally found in bulk [5] and thin film [12] superconductors with weak pinning corroborates the result [13] according to which the mean field approximation [1,2] is not valid just for the infinite homogeneous space. The experimental results [12] show that the prediction [1,2] is not valid at least for two-dimensional superconductor. A mixed state without long-range phase coherence, but no Abrikosov state, is observed down to $H_{c4} \approx 0.005H_{c2}$ in a film with weak pinning [12]. According to [14], the long-range order cannot be realized also in three-dimensional superconductor. The observation of the Abrikosov state below $H_{c4} \approx 0.98H_{c2}$ [5] in bulk superconductor with weak pinning does not refute this result since the real sample has finite sizes. Thus, the experimental results and fluctuation theory are indicative of the non-validity of [1,2] for the ideal case and crystalline long-range order of vortex lattice cannot be realized in real samples [3].

## ACKNOWLEDGMENTS

The work was supported by RFBR, Grant 04-02-17068. I thank LT24 Financial Support Committee for the offer of financial assistance to attend the LT24 Conference.